\documentclass[aps,prd,10pt,amsmath,amssymb,reprint,nofootinbib,groupedaddress,superscriptaddress,floatfix]{revtex4-2}

\usepackage{enumerate}
\usepackage{graphicx}
\usepackage{dcolumn}
\usepackage{bm}
\usepackage{microtype}
\usepackage{xcolor}
\definecolor{navyblue}{rgb}{0.0, 0.0, 0.5}
\usepackage{hyperref}
\hypersetup{
    colorlinks=true,
    linktoc=all,
    allcolors=navyblue
}

\usepackage[capitalize]{cleveref}
\crefname{section}{Sec.}{Secs.}
\Crefname{section}{Sec.}{Secs.}
\crefname{appendix}{App.}{Apps.}
\Crefname{appendix}{App.}{Apps.}
\Crefname{figure}{Fig.}{Figs.}
\crefname{figure}{Fig.}{Figs.}
\creflabelformat{equation}{#2#1#3}

\usepackage{color,environ}
\usepackage{fontawesome5}
\definecolor{orcidlogocol}{rgb}{0.65, 0.807, 0.223}
\newcommand{\orcid}[1]{$\,$\href{https://orcid.org/#1}{\textcolor{orcidlogocol}{\footnotesize\faOrcid}}}
\usepackage{etaremune}
\usepackage{enumitem}

\graphicspath{{./}{./}}

\newcommand*\p[1]{\left(#1\right)}

\newcommand*\f[2]{\frac{#1}{#2}}

\newcommand{\calL}{\mathcal{L}}
\newcommand{\calO}{\mathcal{O}}

\newcommand{\Mpl}{M_\mathrm{pl}}
\newcommand{\Heq}{H_\mathrm{eq}}

\newcommand{\LQCD}{\Lambda_\text{QCD}}

\NewEnviron{eq}{%
\begin{align}\begin{split}
  \BODY
\end{split}\end{align}
}

\usepackage[normalem]{ulem}
\usepackage{mathtools} 

\makeatletter
\newcommand{\amarki}{\faCoffee}
\newcommand{\amarkii}{\faRebel}
\newcommand{\amarkiii}{\faPaperPlane[regular]}
\newcommand{\amarkiv}{\faBeer}
\def\@fnsymbol#1{{\ifcase#1\or \amarki\or \amarkii\or \amarkiii\or \amarkiv \else\@ctrerr\fi}}
\makeatother

\begin{document}
\raggedbottom

\title{Scalar relics from the hot Big Bang}

\author{David Cyncynates\orcid{0000-0002-2660-8407}}
\email{davidcyn@uw.edu}
\affiliation{Department of Physics, University of Washington, Seattle, WA 98195, U.S.A.}

\author{Olivier Simon\orcid{0000-0003-2718-2927}}
\email{osimon@princeton.edu}
\affiliation{Princeton Center for Theoretical Science, Princeton University, Princeton, NJ 08544, U.S.A.}

\date{\today}
\begin{abstract}
In this \textit{Letter}, we motivate the fact that couplings between a scalar field and the Standard Model with strengths $10^{-6}(m_\phi/{\rm eV})^{-1/4}$ relative to gravity yield the total measured cosmological dark matter abundance over a broad mass range of $10^{-12}$ to $10^{14}\ \rm{eV}$. Remarkably, this result holds with minimal sensitivity to whether the scalar couples to electrons, photons, hadrons, or other particles at laboratory energy scales, thereby linking fifth force experiments to the search for dark matter.
\end{abstract}

\maketitle

The prototypical example of a fifth force is described by the addition of a massive singlet scalar to the Standard Model (SM), whose exchange among SM particles leads to the emergence of a Yukawa potential between them. Because a singlet scalar is not associated with any conserved charges, it can couple to any SM operator, generically mediating forces between all known particles. However, the energy scales involved in laboratory tests for fifth forces and those of the early Universe can differ by many orders of magnitude.
In this companion \textit{Letter} to Ref.~\cite{Cyncynates:2024bxw}, we argue that the SM generically generates a cosmological abundance of the scalar field, nearly independent of how the coupling manifests in the laboratory due to 
the running of gauge couplings, dimensional transmutation, and spontaneous symmetry breaking, 
implying a sharp target for fifth force experiments and dark matter haloscopes.

Consider a real scalar field $\phi$ with Lagrangian
\begin{eq}
\label{eq:lagrangian}
    \mathcal L_\phi = \frac{1}{2}\partial_\mu \phi \partial^\mu \phi - \f12 m_\phi^2\phi^2 + \calL_{\rm int.},
\end{eq}
where $m_\phi$ is the mass of $\phi$, and $\calL_{\rm int.}$ contains interactions of $\phi$ with SM particles. The dimensionless scalar field $\varphi = \sqrt{4\pi G}\phi$, where $G$ is Newton's constant, characterizes $\phi$ relative to gravitational potentials. Because $\phi$ has no internal symmetries, it can multiply any SM operator, so variation in its classical expectation value effectively generates space and time dependence of the SM parameters (i.e.~the fundamental constants of the SM). The field-dependence of a generic SM parameter $\zeta$ is then described by the scalar response function
$d'_\zeta(\varphi) \equiv \zeta(\varphi)^{-1}(d\zeta(\varphi)/{d\varphi})$ \cite{Damour:2010rp}.
We consider only the leading interactions of $\phi$, in which case $\zeta[\varphi]= \zeta[0](1 + d_\zeta^{(1)}\varphi)$, where $d_\zeta^{(1)} \approx 1$ corresponds to a gravitational strength interaction.

In the early Universe, the SM plasma generates an effective potential for the scalar $V_{\rm eff}(\phi)$ whose minimum is displaced relative to $\phi$'s vacuum expectation value \cite{Damour:1994zq,Batell:2021ofv,Batell:2022qvr,Gan:2023wnp,Alachkar:2024crj}.$^1$ In a homogeneous Universe with fixed parameters $\{\zeta\}$, the thermal path integral evaluated over the SM species at finite temperature $T$ corresponds to the thermodynamic pressure of the SM sector. Thus, the effective potential for $\phi$ becomes the $\phi$-dependent pressure of the system, where the fundamental constants acquire their local $\phi$-dependent values \cite{Dolan:1973qd,Weinberg:1974hy}:
\begin{eq}
V_\text{eff}(\phi) = -P\left(\{\zeta (\varphi)\}\right).
\end{eq}
At high temperature, the leading contribution to the pressure is independent of the fundamental constants. Therefore, the force on $\phi$ is driven by the largest subleading terms, primarily controlled by the subset of dimensionless SM couplings $\{g_i\}$. These terms capture the interactions between various particles in the plasma, and we denote this pressure of interactions $P_{\rm int.}\sim \sum_i g_i^2(\varphi) T^4$ \cite{Buchmuller:2003is,Buchmuller:2004xr,Lillard:2018zts}.

\begin{figure}
    \centering
    \includegraphics[width=\columnwidth]{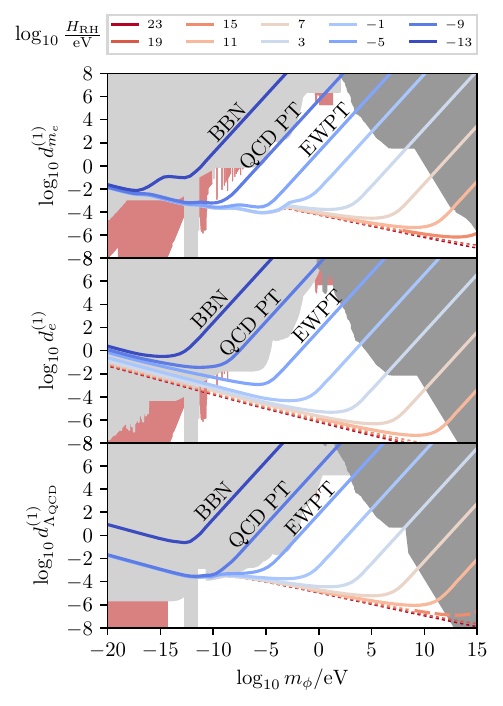}
    \caption{The parameter space of scalar couplings that yields the correct relic abundance as a function of the Hubble rate at reheating $H_{\rm RH}\approx T_{\rm RH}^2/\Mpl$. The labels BBN, QCD PT, and EWPT correspond to instantaneous reheating to temperatures just above the onset of BBN, the QCD Phase Transition, and the Electroweak Phase Transition respectively. 
    Dashed lines indicate where the scalar cannot account for all of the dark matter due to isocurvature constraints, while dotted lines mark where misalignment from inflationary fluctuations is expected to dominate over thermal misalignment. For further, subleading considerations, see Figure~6 of Ref.~\cite{Cyncynates:2024bxw}.
    Each panel corresponds to a scalar that couples \emph{only} to one particular atomic constituent at laboratory energies: the electron mass, the electromagnetic coupling constant, and the QCD scale. The particular case of a coupling to the electron mass has a somewhat stronger dependence on the details of the scalar coupling (see the discussion in Ref.~\cite{Cyncynates:2024bxw}) -- for this figure, we assume the scalar couples to the electron mass through the Higgs potential, meaning that every other, heavy fundamental fermion obtains the same coupling. 
    Observational bounds consisting of equivalence principle tests \cite{Hees:2018fpg,Fischbach:1996eq,MICROSCOPE:2022doy,Smith:1999cr,Schlamminger:2007ht} and inverse square law tests \cite{Adelberger:2003zx,KONOPLIV2011401,Mostepanenko:2020lqe,Kapner:2006si,Lee:2020zjt,Yang:2012zzb,Tan:2020vpf,Chen:2014oda,Ke:2021jtj,Hoskins:1985tn,Geraci:2008hb,Sushkov:2011md}, atomic clocks and spectroscopy  \cite{Filzinger:2023zrs,Tretiak:2022ndx,Oswald:2021vtc,Zhang:2022ewz,2022PhRvL.129x1301K,BACON:2020ubh,VanTilburg:2015oza,Aharony:2019iad,Hees:2016gop,Sherrill:2023zah}, resonant cavities and interferometry \cite{Savalle:2020vgz,Vermeulen:2021epa,Aiello:2021wlp,Kennedy:2020bac,Gottel:2024cfj}, resonant bar and mechanical/atomic oscillators \cite{Branca:2016rez,Campbell:2020fvq}, and astrophysical limits \cite{Hardy:2016kme,Bottaro:2023gep,NANOGrav:2023hvm,Baryakhtar:2020gao,Hoof:2024quk,Fiorillo:2025zzx} are in light gray while astrophysical and cosmological bounds on decaying scalar dark matter are in dark gray \cite{Janish:2023kvi,Cadamuro:2011fd,Wadekar:2021qae,Yin:2024lla,Todarello:2023hdk,Grin:2006aw,Carenza:2023qxh,Porras-Bedmar:2024uql}.
    Prospective sensitivity curves for experiments are plotted in red (top) \cite{Badurina:2021rgt,Arvanitaki:2015iga,Arvanitaki:2017nhi,2021QS&T....6d4003A,Antypas:2022asj,Manley:2019vxy} (middle) \cite{Arvanitaki:2017nhi,Badurina:2021rgt,Arvanitaki:2015iga,Hirschel:2023sbx,2021QS&T....6d4003A,Arvanitaki:2014faa,Geraci:2018fax,2021PhRvA.103d3313K,Manley:2019vxy} (bottom) \cite{Arvanitaki:2014faa,Arvanitaki:2017nhi}.}
    \label{fig:Parameter-Space}
\end{figure}

A central feature of the SM is the fact that the constants of Nature flow and mix as a function of the energy scales of the interactions taking place. Hence the numerical values of fundamental constants observed in low-energy laboratory experiments, often taken to be $\{e,m_e,\LQCD,m_u,m_d\}$ \cite{Badurina:2021rgt,Arvanitaki:2015iga,Arvanitaki:2017nhi,2021QS&T....6d4003A,Antypas:2022asj,Manley:2019vxy,Arvanitaki:2017nhi,Badurina:2021rgt,Arvanitaki:2015iga,Hirschel:2023sbx,2021QS&T....6d4003A,Arvanitaki:2014faa,Geraci:2018fax,2021PhRvA.103d3313K,Manley:2019vxy,Arvanitaki:2014faa,Arvanitaki:2017nhi,Hees:2018fpg,Adelberger:2003zx,Fischbach:1996eq,KONOPLIV2011401,MICROSCOPE:2022doy,Hardy:2016kme,Bottaro:2023gep,Branca:2016rez,Filzinger:2023zrs,Tretiak:2022ndx,Savalle:2020vgz,Vermeulen:2021epa,Aiello:2021wlp,Campbell:2020fvq,Oswald:2021vtc,Kennedy:2020bac,Zhang:2022ewz,2022PhRvL.129x1301K,Gottel:2024cfj,NANOGrav:2023hvm,OHare:2020wah,Hees:2018fpg,Adelberger:2003zx,KONOPLIV2011401,Branca:2016rez,BACON:2020ubh,Tretiak:2022ndx,Savalle:2020vgz,VanTilburg:2015oza,Zhang:2022ewz,Aharony:2019iad,Vermeulen:2021epa,Gottel:2024cfj,Aiello:2021wlp,Campbell:2020fvq,Filzinger:2023zrs,Oswald:2021vtc,Hees:2016gop,Kennedy:2020bac,Sherrill:2023zah,Wadekar:2021qae,Cadamuro:2011fd,POLARBEAR:2023ric,OHare:2020wah,Janish:2023kvi,Yin:2024lla,Todarello:2023hdk,Grin:2006aw,Carenza:2023qxh,Porras-Bedmar:2024uql,Mostepanenko:2020lqe,Kapner:2006si,Lee:2020zjt,Smith:1999cr,Schlamminger:2007ht,Yang:2012zzb,Tan:2020vpf,Chen:2014oda,Ke:2021jtj,MICROSCOPE:2022doy,Hoskins:1985tn,Raffelt:2012sp,Hardy:2016kme,Geraci:2008hb,Bottaro:2023gep,Sushkov:2011md,OHare:2020wah} (the strength of electromagnetism, the mass of the electron, the confinement scale of quantum chromodynamics (QCD), and the mass of the up and down quarks, respectively), are different than those at higher energies reached in the early Universe. The associated couplings of the scalar to photons, electrons, gluons and light quarks, correspondingly obtain temperature-dependent values.$^2$

In particular, the dimensionless coupling $g_i$ at one energy scale $\bar\mu$ is a function of all the other fundamental constants at some fixed reference scale $\bar\mu_0$:
\begin{align}\label{eqn:gi}
    g_i[\bar\mu,\varphi] = g_i[\bar\mu,\varphi;\{\zeta_j[{\bar\mu_0,\varphi}]\}],
\end{align}
where $\zeta_j[\bar\mu,\varphi]$ indicates the value of the SM parameter $\zeta_j$ at the energy scale $\bar\mu$ and the background scalar value $\varphi$. 
Differentiating the set of relations \cref{eqn:gi} gives the scale dependence of the response functions, which we truncate at leading order in $\varphi$ \cite{Damour:2010rp}:
\begin{align}
    d_{g_i}^{(1)}(\bar\mu) = \sum_{j = 1} \left.\frac{\partial \log g_i[\bar\mu,\varphi]}{\partial \log\zeta_j[\bar\mu_0,\varphi]}\right|_{\varphi = 0} d_{\zeta_j}^{(1)}(\bar\mu_0).
\end{align}
Setting $\bar\mu = T$ and $\bar\mu_0 = 0$ to match onto the values of the fundamental constants in the early and present Universe respectively, this mixing implies that the leading order temperature dependence of the effective potential is
\begin{align}\label{eqn:eff-pot}
    \f{\partial V_\text{eff}(\varphi)}{\partial\varphi} = -T^4\sum_i\left.\f{\partial P_{\rm int.}(\{g_j[T,\varphi]\})}{T^4\partial\log\zeta_i(0,\varphi)}\right|_{\varphi = 0}d_{\zeta_i}^{(1)}+ \calO(T^2).
\end{align}
A substantial literature exists on precision computations of pressures in the SM {\cite{Arnold:1994eb,kapusta2007finite,Kajantie:2002wa,Gynther:2003za,Gynther:2005av,Gynther:2005dj,Gynther:2006wq,Laine:2015kra}}. As we are concerned with small variations of the fundamental parameters, we truncate these results at leading order in the SM parameters.

From these formulae, it is clear that the various $T^4$ contributions to $V_{\rm eff}(\phi)$ depend on all the SM parameters as measured at some experimentally relevant reference scale, and that the dependence of any mass scale or interaction strength on $\phi$ generates such a term. For simplicity, we assume that the Universe reheats instantaneously and that the SM bath achieves a maximum temperature $T_{\rm RH}$.$^3$ The scalar field then acquires an irreducible non-thermal relic abundance from the changing effective potential:
\begin{eq}
\label{eq:misalignment_gauge_coupling}
\f{\Omega_{\varphi}}{\Omega_{\rm DM}} \propto&  \left(\sum_i\left.\f{\partial P_{\rm int.}(\{g_j[T_{\rm RH},\varphi]\})}{T_{\rm RH}^4\partial\log\zeta_i(0,\varphi)}\right|_{\varphi = 0}d_{\zeta_i}^{(1)}\right)^2 \times\\&\sqrt{\f{m_\phi}{\Heq}}\times
\begin{cases}
\log^2\p{\frac{T^2_{\rm RH}}{\Mpl m_\phi}}, & m_\phi \lesssim \frac{T_{\rm RH}^2}{\Mpl}, \\
\left(\frac{T^2_{\rm RH}}{\Mpl m_\phi}\right)^{5/2},& m_\phi \gtrsim \frac{T_{\rm RH}^2}{\Mpl},
\end{cases}
\end{eq}
where $\Omega_{\varphi,\rm DM}$ are the fraction of the critical density in the scalar and dark matter respectively. The contribution of other processes to the scalar relic abundance is estimated in Ref.~\cite{Cyncynates:2024bxw}, including thermal scattering and inflationary fluctuations, and it is shown that they can be consistently neglected over much of the available parameter space. As the interaction pressure is determined only by dimensionless couplings at leading order, the first line of \cref{eq:misalignment_gauge_coupling} simplifies to the sum over products of $d_{\zeta_i}^{(1)}$ and the corresponding dimensionless couplings. In the case that the dimensionless couplings are ${\cal O}(1)$ and $m_\phi < H_{\rm RH}$, \cref{eq:misalignment_gauge_coupling} can be inverted:
\begin{align}\label{eq:prior}
d_\zeta^{(1)} \sim 10^{-6}(m_\phi/\text{eV})^{-1/4}.
\end{align}

The sense in which this sets a prior on parameter space can be clarified by considering the putative discovery of a laboratory fifth force with a coupling to matter exceeding \cref{eq:prior}. Such a large coupling, given the measured abundance of cosmological dark matter, implies that the relic abundance is suppressed by a cancellation between terms in the bracketed sum in \cref{eq:misalignment_gauge_coupling}. 
For $T_\text{RH}$ greater than the electroweak scale, this can only happen if the scalar couples to the standard model exclusively through a superrenormalizable coupling to the Higgs, where $\varphi$ modulates only the Higgs mass parameter above the electroweak scale~\cite{Shtanov:2021uif,Shtanov:2022xew,Batell:2022qvr}. Even in this special case, our conclusions are largely unchanged: one simply cuts off the thermal effective potential at the electroweak scale, where the cancellation occurs. Further, the ratios of the scalar's couplings at laboratory scales are fixed by renormalization group running from the electroweak scale: $d_{\LQCD}^{(1)} = -(6/27)\, d_{m_f}^{(1)}$ and $d_{e}^{(1)} = -(16\alpha_{\rm EM}/3\pi)\, d_{m_f}^{(1)}$, where $m_f$ are fermion masses. 

More generally, when the pressure of interactions is suppressed by a precise cancellation, this typically happens only at a single temperature. Differential running of masses and couplings prevents such a cancellation from persisting across scales. As a result, couplings to heavy particles not accessible in laboratory tests will usually contribute additional pressure, increasing the relic abundance of $\phi$, implying that the typical allowed couplings are smaller than \cref{eq:prior}. 

We illustrate the coupling strengths which yield the correct dark matter relic abundance in \cref{fig:Parameter-Space}, scanning over the allowable reheat temperatures, and under the assumption that the scalar only modulates only one of the electron mass $[d_{m_e}^{(1)}]$, the electromagnetic coupling constant $[d_{e}^{(1)}]$, or the QCD scale $[d_{\LQCD}^{(1)}]$ ($\sim$baryon masses) at laboratory scales$^4$ --- an assumption often \cite{Badurina:2021rgt,Arvanitaki:2015iga,Arvanitaki:2017nhi,2021QS&T....6d4003A,Antypas:2022asj,Manley:2019vxy,Arvanitaki:2017nhi,Badurina:2021rgt,Arvanitaki:2015iga,Hirschel:2023sbx,2021QS&T....6d4003A,Arvanitaki:2014faa,Geraci:2018fax,2021PhRvA.103d3313K,Manley:2019vxy,Arvanitaki:2014faa,Arvanitaki:2017nhi,Hees:2018fpg,Adelberger:2003zx,Fischbach:1996eq,KONOPLIV2011401,MICROSCOPE:2022doy,Hardy:2016kme,Bottaro:2023gep,Branca:2016rez,Filzinger:2023zrs,Tretiak:2022ndx,Savalle:2020vgz,Vermeulen:2021epa,Aiello:2021wlp,Campbell:2020fvq,Kennedy:2020bac,Zhang:2022ewz,2022PhRvL.129x1301K,Gottel:2024cfj,NANOGrav:2023hvm,OHare:2020wah,Hees:2018fpg,Adelberger:2003zx,KONOPLIV2011401,Branca:2016rez,BACON:2020ubh,Tretiak:2022ndx,Savalle:2020vgz,VanTilburg:2015oza,Zhang:2022ewz,Aharony:2019iad,Vermeulen:2021epa,Gottel:2024cfj,Aiello:2021wlp,Campbell:2020fvq,Filzinger:2023zrs,Hees:2016gop,Kennedy:2020bac,Sherrill:2023zah,Wadekar:2021qae,Cadamuro:2011fd,POLARBEAR:2023ric,OHare:2020wah,Janish:2023kvi,Yin:2024lla,Todarello:2023hdk,Grin:2006aw,Carenza:2023qxh,Porras-Bedmar:2024uql,Mostepanenko:2020lqe,Kapner:2006si,Lee:2020zjt,Smith:1999cr,Schlamminger:2007ht,Yang:2012zzb,Tan:2020vpf,Chen:2014oda,Ke:2021jtj,MICROSCOPE:2022doy,Hoskins:1985tn,Raffelt:2012sp,Hardy:2016kme,Geraci:2008hb,Bottaro:2023gep,Sushkov:2011md,OHare:2020wah} (though not always \cite{Oswald:2021vtc,Banerjee:2022sqg}) made when displaying the sensitivity of experimental searches. The similarity of each panel in the figure shows that a scalar mediating a fifth force couples to the early Universe plasma in a way largely independent of its low-energy interactions. Scalars with mass smaller than the Hubble rate at reheating $T_{\rm RH}^2/\Mpl$ are insensitive to the reheat temperature, in contrast to scalars with larger mass; scalar couplings comparable to $d_\zeta^{(1)} \sim 10^{-6}(m_\phi/\text{eV})^{-1/4}$ yield a relic abundance that matches the dark matter density so long as $m_\phi \lesssim T_{\rm RH}^2/\Mpl$. 
Low reheat temperature can yield deviations from the from the universal scaling \cref{eq:prior}. 
The diversity of scalar behavior at low reheat temperature can be understood by comparing the low $H_{\rm RH}$ (blue) curves among the three panels of \cref{fig:Parameter-Space}. 
Planned experiments will probe these couplings \cite{Badurina:2021rgt, Arvanitaki:2015iga}, particularly around scalar masses of $10^{-12} \text{eV}$ where current astrophysical constraints are weak and sensitivity to the reheat temperature is minimal.

\acknowledgements
We thank Masha Baryakhtar, Anson Hook, Junwu Huang, Mikko Laine, Ken Van Tilburg, Neal Weiner, Zachary Weiner, and Lawrence Yaffe for helpful discussions and correspondences. We thank Jedidiah Thompson for collaboration during the early stages of this project. We thank Saarik Kalia for insightful comments on the draft. This work was completed in part at the Perimeter Institute for Theoretical Physics. Research at Perimeter Institute is supported in part by the Government of Canada through the Department of Innovation, Science and Economic Development Canada and by the Province of Ontario through the Ministry of Colleges and Universities. D.C.\ is supported through the Department of Physics and College of Arts and Science at the University of Washington and by the U.S. Department of Energy Office of Science under Award Number DE-SC0024375. Experimental limits and prospects presented in this paper were in part compiled using the GitHub repository associated with Ref.~\cite{AxionLimits}.

\section*{Endnotes}
\noindent\footnotesize
$^1${There is also a zero-temperature contribution: we estimate its size and illustrate the parameter space where it can be consistently neglected in Ref.~\cite{Cyncynates:2024bxw}. A conservative naturalness criterion would be that the UV cutoff of SM loops is above the reheat temperature and small enough that the mass is not strongly renormalized: $m_\phi^2 \gtrsim (d_\zeta^{(1)} \zeta)^2 \Lambda_{\rm UV}^4/\Mpl^2\gtrsim (d_\zeta^{(1)} \zeta)^2 H_{\rm RH}^2$.}\\
$^2${In contrast to Ref.~\cite{Damour:2010rp} and much of the literature after it, which uses the notation $d^{(1)}_e$ and $d^{(1)}_g$, these coupling functions are written in our notation as $d^{(1)}_{\alpha_\text{EM}}$ and $d^{(1)}_{\Lambda_\text{QCD}}$, respectively.}\\
$^3${Our results are insensitive to the reheating history provided $m_\phi < H_{\rm RH}$; earlier dynamics are effectively instantaneous from the scalar’s perspective. The study of heavier scalars with different reheat histories is left to future work.}\\
$^4${We treat the light quarks as massless to avoid the technical difficulties associated with their running across the QCD phase transition, which would detract from our main analysis while having minimal quantitative impact.}

\bibliography{bibliography}

\end{document}